\documentclass[letterpaper,journal]{IEEEtran}
\usepackage{amsmath,amsfonts}
\usepackage{algorithmic}
\usepackage{array}
\usepackage[caption=false,font=normalsize,labelfont=sf,textfont=sf]{subfig}
\usepackage{textcomp}
\usepackage{stfloats}
\usepackage{url}
\usepackage{graphicx}
\usepackage{siunitx}
\usepackage{hyperref}
\DeclareSIUnit \rperminute {rpm}
\DeclareSIUnit \belm {Bm}
\begin{document}
\title{Optimizing Cladding Elasticity to Enhance Sensitivity in Silicon Photonic Ultrasound Sensors}
\author{R. Tufan Erdogan, Georgy A. Filonenko, Stephen J. Picken, Peter G. Steeneken, Wouter J. Westerveld
\thanks{Manuscript received xxx; revised xxx; accepted xxx. Date of publication xxx. \\ 
R. T. Erdogan, P. G. Steeneken, and W. J. Westerveld are with the Department of Precision and Microsystems Engineering, Delft University of Technology, Mekelweg 2, 2628 CD Delft, The Netherlands (e-mails: r.t.erdogan@tuldelft.nl, p.g.steeneken@tudelft.nl, w.j.westerveld@tudelft.nl).\\
G. A. Filonenko is with the Department of Materials Science and Engineering, Delft University of Technology, Mekelweg 2, 2628 CD Delft, The Netherlands (e-mail: g.a.filonenko@tudelft.nl).\\
S. J. Picken is with the Advanced Soft Matter, Delft University of Technology, Van der Maasweg 9, 2629 HZ Delft, The Netherlands (e-mail: s.j.picken@tudelft.nl).

© 20XX IEEE.  Personal use of this material is permitted.  Permission from IEEE must be obtained for all other uses, in any current or future media, including reprinting/republishing this material for advertising or promotional purposes, creating new collective works, for resale or redistribution to servers or lists, or reuse of any copyrighted component of this work in other works.
}}
\markboth{XXXX,~Vol.~XXXX, No.~XXXX, XXXX~2024}%
{}
\maketitle
\begin{abstract}
Ultrasound is widely used in medical imaging, and emerging photo-acoustic imaging is crucial for disease diagnosis. Currently, high-end photo-acoustic imaging systems rely on piezo-electric materials for detecting ultrasound waves, which come with sensitivity, noise, and bandwidth limitations. Advanced applications demand a large matrix of broadband, high-resolution, and scalable ultrasound sensors. Silicon photonic circuits have been introduced to meet these requirements by detecting ultrasound-induced deformation and stress in silicon waveguides. Although higher sensitivities could facilitate the exploration of new applications, the high stiffness of the waveguide materials constrains the intrinsic sensitivity of the silicon photonic circuits to ultrasound signals. Here, we explore the impact of the mechanical properties of a polymer cladding on the sensitivity of silicon photonic ultrasound sensors. Our model and experiments reveal that optimizing the polymer cladding's stiffness enhances the resonance wavelength sensitivity. Experimentally, we show a fourfold increase in the sensitivity compared to the sensors without a cladding polymer and, a twofold sensitivity increase compared to the sensors with a cladding polymer of saturated cross-linking density.  Interestingly, comparing experiments with the optomechanical model suggests that the change in Young's Modulus alone cannot explain the sensitivity increase. In conclusion, polymer-coated silicon photonic ultrasound sensors exhibit potential for advanced photo-acoustic imaging applications. It offers the prospect of increasing the ultrasound detection sensitivity of silicon photonic ultrasound sensors while using CMOS-compatible processes. This paves the way to integrate the polymer-coated silicon photonic ultrasound sensors with electronics to utilize the sensors in advanced medical imaging applications. 

\end{abstract}

\begin{IEEEkeywords}
 ultrasound, silicon photonics, sensors, PDMS, photo-elasticity
\end{IEEEkeywords}

\section{Introduction}
\IEEEPARstart{P}{recisely} identifying skin diseases and determining their location under the skin is important for medical applications. Conventional medical ultrasound imaging is widely used to image tissue under the skin. Recently, photoacoustic imaging (PAI) has emerged as a new and promising medical imaging technique \cite{riksen_photoacoustic_2023,xia_photoacoustic_2014,xia_whole-body_2012} that is paving its way to becoming a mainstream method in clinical settings \cite{kruger_dedicated_2013,dahlstrand_spectral_2020,van_der_sanden_vascular_2020}. PAI provides rapid, functional, and high-resolution imaging deep into the tissue. Unlike traditional pulse-echo ultrasound imaging, PAI involves directing a short pulse of light into the body, where it is absorbed by the tissue. This absorption causes the tissue to heat up and expand, thereby generating ultrasound waves due to the optical absorption contrast. The ultrasound waves are then captured by an array of sensors placed on the skin, allowing imaging up to depths of several millimeters \cite{steinberg_photoacoustic_2019} or even centimeters by taking advantage of the low scattering of ultrasound waves in tissue \cite{beard_biomedical_2011}. To create a tomographic image using this method, it is necessary to utilize sensor arrays that simultaneously receive the ultrasonic wavefield at multiple locations.

In PAI applications, various types of ultrasound sensors are used to create an image in an array or matrix formation, including piezo-electric transducers \cite{yuan_real-time_2013}, capacitive micromachined ultrasound transducers (CMUTs) \cite{oralkan_capacitive_2002}, and optical ultrasound detectors \cite{fu_optical_2022}. Due to noise, size, and bandwidth limitations of piezoelectric and CMUT sensors \cite{westerveld_optical_2019}, interest in using optical ultrasound sensors for PAI applications is growing. To address this, silicon photonic microchips are deployed as a promising optical ultrasound sensing technology \cite{westerveld_optical_2019,westerveld_sensitive_2021,rosenthal_embedded_2014,zarkos_ring_2019,shnaiderman_submicrometre_2020}. 

Silicon photonic microchips can be fabricated in existing complementary metal oxide semiconductor (CMOS) foundries \cite{zarkos_fully_2023}. This enables us to scale the sensor's production with diminishing costs at high volumes. Moreover, high refractive index contrast between the silicon core and cladding materials of the waveguides enables us to design sensor dimensions smaller than \qty{1}{\micro\meter} without compromising the sensitivity, bandwidth, and noise characteristics \cite{hazan_silicon-photonics_2022}. Thus, silicon photonic ultrasound sensors can meet the requirements of advanced PAI applications in size, bandwidth, and scalability when designed as ultrasound sensors. 

The silicon photonic ultrasound sensors detect sound pressure waves via the deformation of optical waveguides due to ultrasound pressure. This leads to a change in either the length of an optically resonant circuit element \cite{westerveld_optical_2019,leinders_sensitive_2015} or a change in the effective refractive index of the propagating mode due to geometrical and photo-elastic effects \cite{tsesses_modeling_2017,schumacher_PSPDMS}. The latter effect allows designers to utilize the scalability of silicon photonics using highly elasto-optic materials as a cladding to enhance ultrasound detection sensitivity \cite{ravi_kumar_enhanced_2019} without utilizing complex fabrication methods. Recent demonstrations of PAI with polymer-coated silicon photonic ultrasound sensors \cite{hazan_silicon-photonics_2022,nagli_silicon_2023,harary_all-optical_2023,hazan_miniaturized_2022,ding_broadband_2022} have shown the enhancement of the sensitivity of silicon photonic ultrasound sensors in comparison to silicon-dioxide (SiO$_{2}$) cladding or water cladding. In addition to this, silicon photonic ultrasound sensors clad with a polydimethylsiloxane (PDMS) polymer layer \cite{hazan_silicon-photonics_2022} have recently been shown to be effective in the rejection of surface acoustic waves, which otherwise increase distortions and artifacts in ultrasound images.  Nevertheless, further sensitivity improvements are needed to enable new applications with silicon photonic ultrasound sensing \cite{westerveld_sensitive_2021}.

In this work, we study the effect of mechanical and optomechanical properties of polymer claddings on the sensitivity of silicon photonic ultrasound sensors. We aim to enhance the sensitivity of polymer-cladded silicon photonic ultrasound sensors by implementing tailored polymer coatings. First, we model the refractive index sensitivity of these types of waveguides, i.e., the change in effective refractive index $n_e$, due to the incident ultrasound plane-wave with pressure $P$. We find that refractive index sensitivity scales inversely with cladding material stiffness since lower stiffness leads to larger deformations. Second, to validate the model, we experimentally investigate the sensor sensitivity as a function of cladding stiffness. We prepare sensors with PDMS cladding material of different stiffness and measure the dependence of their refractive index on applied ultrasound pressure. We show that optimizing the PDMS cladding stiffness can increase the pressure-induced shift in resonance wavelength by a factor of four compared to sensors without a cladding layer and by a factor of two compared to sensors with a cladding of standard PDMS stiffness. Our results directly apply to silicon photonic ultrasound sensors where the refractive index change of the cladding material is used as an acousto-optic transduction mechanism.
\begin{figure}[!t]
\centering
\includegraphics{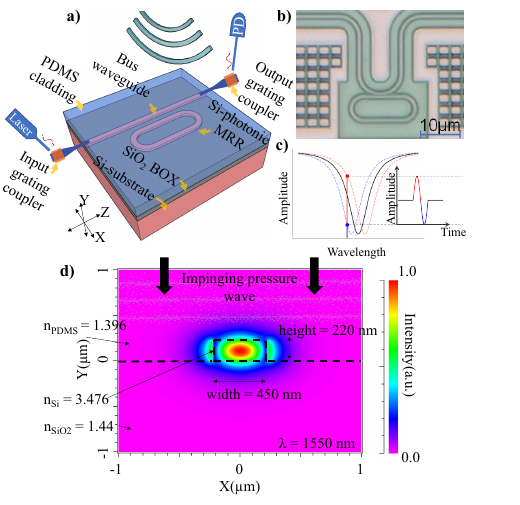}
\caption{Silicon photonic ultrasound sensor with PDMS cladding. a) 3D sketch of PDMS cladded silicon photonic ultrasound sensor and schematic of the optical path. PD: photodetector. b) Optical microscope image of the micro-ring resonator (MRR) after PDMS spin-coating. c) Schematic of the read-out method by transmission output intensity monitoring when the laser wavelength is set to flank wavelength on the resonance of the MRR. d) Cross-sectional sketch of the waveguide (dashed black lines) overlaid on the mode field simulation. The simulation result shows the electric field intensity distribution in the PDMS cladding and the silicon waveguide core. Refractive index sensing occurs due to the intersection of the evanescent field with the cladding material.}
\label{conceptFig}
\end{figure}
\section{Results}
\subsection{Polymer-cladded silicon photonic ultrasound sensors}
\label{sec:sensor_def}
\noindent Fig. \ref{conceptFig}(a) shows a 3D sketch of the silicon photonic ultrasound sensor used in this study. The sensor comprises a bus waveguide and a micro-ring resonator covered with PDMS cladding. A tunable laser is used to interrogate the ring resonator. Input light from the laser is launched with a single-mode fiber. A TE mode grating coupler is utilized to couple the light from the single-mode fiber to the bus waveguide. Similarly, another copy of the grating coupler is used at the output to couple the transmitted light from the bus waveguide to a multi-mode fiber. The light intensity is monitored at the output fiber by a photodetector.

The silicon photonic waveguide that defines both the micro-ring resonator and the bus waveguide in  Fig. \ref{conceptFig}(b) has a width of \qty{450}{\nm} and a height of \qty{220}{\nm}. The bus waveguide comprises two \ang{90} circular bends and a straight waveguide between the bends. The length of the straight portion of the coupling section is \qty{500}{\nm}. The micro-ring resonator comprises four \ang{90} circular bends of \qty{3}{\micro\metre} radius and two identical straight sections with a \qty{10}{\micro\metre} length. The directional coupling gap between the ring resonator waveguide and the bus waveguide is \qty{200}{\nano\metre}.  Devices with the same nominal design dimensions are used throughout this study. 

Additionally, the PDMS cladding has a thickness of \qty{5}{\micro\metre}. This thickness was analyzed to determine whether it is significantly smaller than the acoustic wavelength within the PDMS layer. Reported values for the speed of sound across different acoustic frequencies and curing formulations of PDMS are available in the literature \cite{PDMS_att}. As a preliminary estimate, we calculated half of the acoustic wavelength in PDMS to be approximately \qty{26.3}{\micro\metre}, using a sound speed of \qty{1028}{\meter/\second} at the central frequency of our transducer, \qty{19.6}{\mega\hertz}. To confirm this estimate, we used the same speed of sound, reported Young's modulus and density values for PDMS \cite{sharfeddin_comparison_2015}, and applied the equation for the speed of longitudinal acoustic waves \cite{elmore_physics_1985,achenbach_wave_1999,lempriere_ultrasound_2002}. This calculation yielded a Poisson's ratio near the expected value of $0.5$. Based on this verification, both acoustic absorption losses and the impact of standing waves within the PDMS layer are negligible, as the acoustic wavelength is considerably larger than the PDMS thickness.

The read-out method employed for the silicon photonic sensor involves monitoring the time-dependent transmitted output intensity, as illustrated schematically in Fig. \ref{conceptFig}(c). The laser wavelength is adjusted to align with the flank of the transmission dip, which is caused by the micro-ring resonance. Any change in the resonance wavelength of the resonator consequently changes the transmitted intensity. To optimize the sensitivity, we tune the laser wavelength to obtain maximum output pulse amplitude and operate the sensor at this wavelength for the ultrasound pulse recordings. The inset in Fig. \ref{conceptFig}(c) qualitatively presents the temporal response of the resonator's output when subjected to a pressure pulse impinging on the resonator. This perturbation induces a transmission shift due to the resonance wavelength alteration.

\begin{figure*}[!t]
\centering
\includegraphics{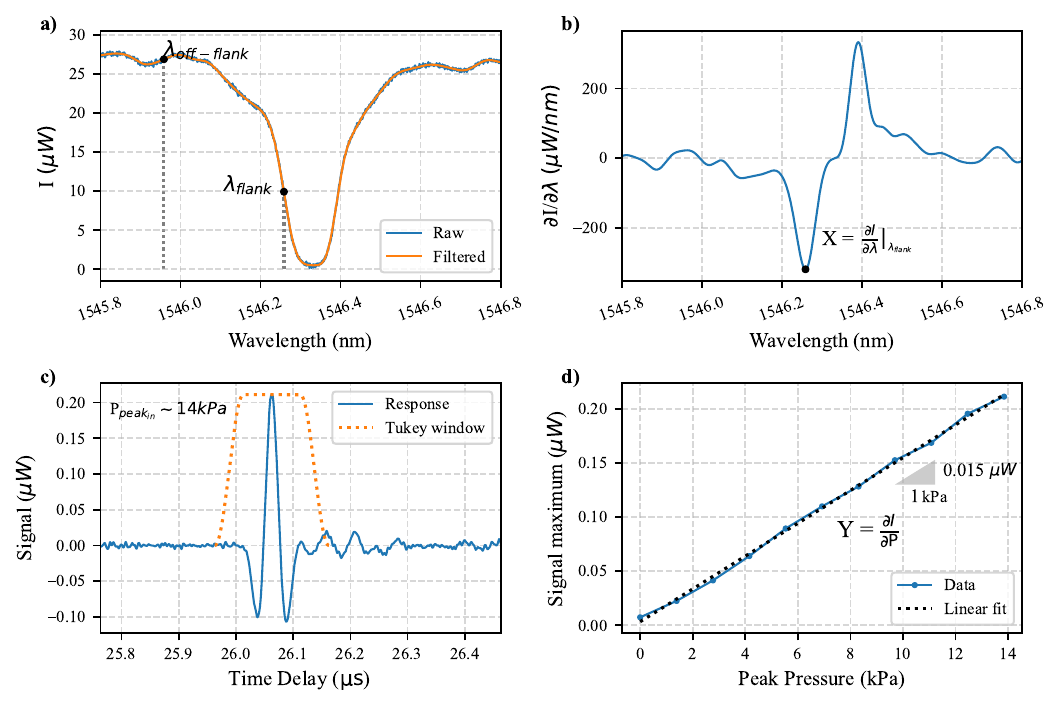}
\caption{Optical and ultrasound characterization of PDMS coated silicon photonic ultrasound sensors. a) Optical transmission spectrum of the ring resonator, measured as the DC output of the photodetector, shows the intensity readings across a sweep of input laser wavelengths. The orange line corresponds to the filtered render of the data intended for derivative computation. b) The derivative of the recorded intensity values in (a). This curve's minimum and maximum points correspond to the flank wavelengths on the left and right flanks of the resonance spectrum.  c) AC time trace (average of 3000 pulse recordings) of the sensor's response to an ultrasound pulse with a peak pressure of \qty{13.85}{\kilo\pascal}. A Tukey window was applied to isolate the initial impingement response of the sensor and remove reflections from the chip's bottom. d) Peak values of the recorded AC signals for different applied ultrasound pressures. The slope of the linear fit is extracted as the intensity sensitivity of the silicon photonic ultrasound sensor to a change in ultrasound pressure.}
\label{characterizationFig}
\end{figure*}
To illustrate the interaction between the ultrasound pressure wave and the propagating electric field in the waveguide, Fig. \ref{conceptFig}(d) presents a mode field simulation that depicts the electric field intensity distribution, along with the waveguide edges and buried oxide surface indicated by black dashed lines. FemSIM algorithm in RSoft from Synopsys Inc. is used to simulate the mode field distribution at a wavelength of \qty{1550}{\nm}. A silicon core with a width and height of \qty{450}{\nm}, \qty{220}{\nm} and with a refractive index of \num{3,476} is placed in the simulation window, which is in the size of \numproduct{4x4} \unit{\micro\meter\squared}. The cladding refractive index on top of the waveguide is assigned as \num{1,396}, representing the PDMS refractive index at \qty{1550}{\nm} wavelength,  and the buried oxide refractive index is defined as \num{1,44}. According to this simulation, the total electric field of $TE_{0}$ mode is confined in the waveguide core, but some portion extends beyond its edge as an evanescent field. This evanescent field interacts with the cladding material through elasto-optic effects, and as a result, the effective refractive index of the waveguide mode depends on the refractive index and shape of the cladding layer. Since the evanescent field does not extend beyond \qty{1}{\micro\metre} in the PDMS cladding, the changes in the thickness of the cladding PDMS can not interfere with the propagating mode. Thus, the current geometry is expected to be sensitive to elasto-optic refractive index changes in the cladding material, but not to the geometrical effects of the cladding layer due to acoustic pressure.

The sensitivity of the sensor, $S_{\lambda_{res}}$, is defined as the change in optical resonance wavelength per applied pressure. It can be calculated as
\begin{equation}
\label{eq001}
 	{S_{{\lambda _{res}}}} = \frac{\partial \lambda}{\partial P} = \frac{\frac{\partial I}{\partial P}}{\left.\frac{\partial I}{\partial \lambda}\right\vert_{\lambda_{flank}}}
\end{equation} 
where $\lambda$ is the resonance wavelength, $P$ is the ultrasound pressure, $I$ is the measured intensity of the light.

The optical response of the resonator is plotted in Fig. \ref{characterizationFig}(a), and the slope of this curve is calculated as $\frac{\partial I}{\partial \lambda}$, which is plotted in Fig. \ref{characterizationFig}(b). The minimum of this plot is indicated as $X$ in Fig. \ref{characterizationFig}(b) and used for the calculation. Then, the ultrasound response of the sensor is measured while keeping the wavelength at the flank wavelength ($\lambda_{flank}$) as in Fig. \ref{characterizationFig}(c). The numerator of the last term in Eq. \ref{eq001} is measured by calculating the change in the ultrasound response for different ultrasound pulse amplitudes. The peak response values of these measurements are plotted in Fig. \ref{characterizationFig}(d). The slope of this curve corresponds to $\frac{\partial I}{\partial P}$ (see $Y$ in Fig. \ref{characterizationFig}(d)). Finally, Eq. \ref{eq001} is used to obtain $S_{\lambda_{res}}$.

\subsection{Ultrasound characterization of the sensor}
\label{sec:ultrasound_char}
\begin{figure}[!t]
\centering
\includegraphics{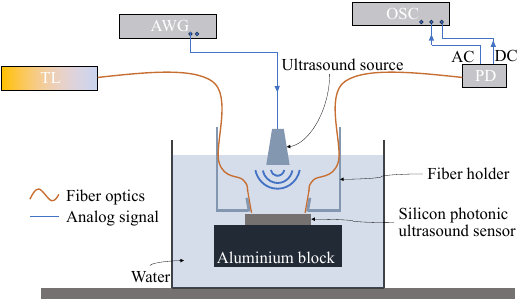}
\caption{Schematic view of ultrasound characterization setup. Fiber holders and ultrasound source transducer are connected to 3-axis positioning stages. AWG is an arbitrary waveform generator, TL is a tuneable laser, OSC is an oscilloscope, and PD is a photodetector.}
\label{setupFig}
\end{figure}

\noindent To characterize the response of the silicon photonic ultrasound sensor, we place the sensor across an ultrasound transducer in a water tank, as shown in Fig. \ref{setupFig}. A Gaussian pulse with a bandwidth of \qty{8}{\mega\hertz}, a center frequency of \qty{18.53}{\mega\hertz}, and a peak-to-peak voltage of \qty{5}{\volt} is generated by using an arbitrary waveform generator (AWG), and applied to the ultrasound transducer. This Gaussian pulse generates a  \qty{13.85}{\kilo\pascal} of pressure amplitude at the sensor's location when calibrated to a hydrophone (see Methods).  The laser wavelength is set to the left flank of the resonance to obtain maximum sensitivity, as annotated in Fig. \ref{characterizationFig}(a,b), and the sensor is aligned below the ultrasound source. Consequently, the response of the sensor is recorded. The recorded response is shown in Fig. \ref{characterizationFig}(c) after averaging over 3000 pulses by an oscilloscope to remove random noise. The pulse generated from the ultrasound transducer arrives at the sensor after a time delay of \qty{26}{\micro\second}. This directly translates to a \qty{38}{\mm} distance between the sensor and the ultrasound transducer. In all experiments, we adjust this distance to the same value by measuring the time delay of pulse arrival.

A signal is recorded for each pressure amplitude as shown in Fig. \ref{characterizationFig}(c), and the peak value of each recorded signal is plotted against the applied pressure in Fig. \ref{characterizationFig}(d), where the pressure is calculated based on the calibrated hydrophone measurement (see Methods). This curve's slope gives the sensor's intensity sensitivity at a given laser input power (\qty{-2}{\deci\belm}), and the slope is denoted as $Y$ in Fig. \ref{characterizationFig}(d). According to the calibration measurements for each pressure amplitude, the peak pressure generated by the transducer is in linear relation with the applied voltage to the transducer. Thus, the linear relation observed in Fig. \ref{characterizationFig}(d) reflects the linearity of the sensor response in the measured pressure interval.

\subsection{Cladding materials}
\noindent The Methods section of this paper proposes a derivation of the theoretical elasto-optic model for silicon photonic micro-ring ultrasound sensors with polymer-cladded waveguides. As a result of this elasto-optic model, we find the sensitivity of these sensors as
\begin{equation}
\label{eq1}
 	{S_{{\lambda _{res}}}} = \frac{{{\lambda _r}}}{{{n_g}}}  \frac{{\partial {n_e}}}{{\partial {n_c}}}  \left[ { - \frac{1}{2}n_c^3{p_{12}}\frac{1}{E}\left( {\frac{{\left( {1 - 2\nu} \right)\left( {1 + \nu} \right)}}{{1 - \nu}}} \right)} \right]
\end{equation} 
where $S_{{\lambda _{res}}}$ is the resonance wavelength sensitivity, $\lambda _r$ is the resonance wavelength, $n_g$ and $n_e$ are the group index and effective index of the waveguide mode, $n_c$ is the cladding material refractive index, $p_{12}$ is the elasto-optic coefficient \cite{Yariv_Yeh_1984}, $E$ is the Young's modulus of the cladding material, and $\nu$ is the Poisson ratio of the cladding material. This result from our theoretical derivation suggests that $S_{{\lambda _{res}}}$ scales inversely with the Young's modulus of the cladding material. 

We examine the influence of cladding stiffness on the sensitivity of resonance wavelength, denoted as $S_{\lambda _{res}}$, by adjusting the stiffness of the PDMS cladding layer. PDMS provides a favorable environment for stiffness modification through polymerization, achieved by mixing a curing agent and a base polymer at a predefined ratio, referred to as the curing ratio. A reduction in the curing agent content in the mixture leads to a decrease in the cross-linked polymer chain density in the cured polymer. Ultimately, it allows us to prepare samples with various Young's Moduli. 

For systematic stiffness modification, PDMS samples with varying curing ratios were prepared. These mixtures were spin-coated onto silicon photonic chips without a top cladding layer and cured under identical conditions. The chosen curing ratios range from 1:10 to 1:50, with the highest ratio at 1:10. This highest mixing ratio is selected because it is the manufacturer's recommendation for the saturated cross-link density in the polymer chain of the PDMS. Hence, the other samples with a less curing agent in the mixture will be under-cross-linked compared to the saturated ratio. Using this method, the cross-link density is accurately controlled by the mixing-ratio and does not critically depend on curing time or temperature, provided that the PDMS is fully cured. 

\begin{figure}[!t]
\centering
\includegraphics{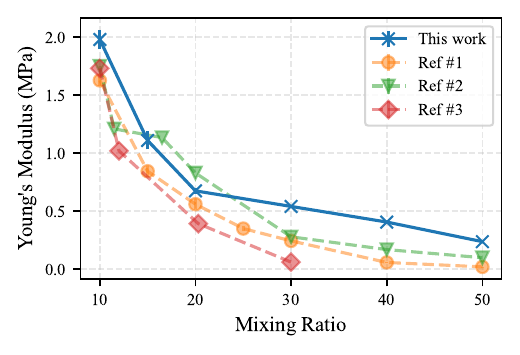}
\caption{Young's modulus of the cladding polymer coated on silicon photonic ultrasound sensors in this study. Results from different studies for the same material under similar conditions are also depicted. Ref. $\# 1$: \cite{glover_extracting_2020}, Ref. $\# 2$: \cite{sharfeddin_comparison_2015}, Ref. $\# 3$: \cite{cho_formulation_2021}}
\label{indentFig}
\end{figure}

The Young's modulus of each sample with a different cross-linking density of the PDMS layer is characterized by using the nano-indentation method. We used a commercial nano-indentation instrument (Piuma, Optics11). Following the ultrasound characterization experiments, we performed nano-indentation at 18 different locations, including the location of the micro-ring resonator used in the ultrasound experiments. The force-displacement curves obtained by nano-indentation are used to calculate the Young's modulus of the sample by JKR fitting of the unloading section of the indentation curve. This fit is performed by the commercial software of the nano-indentation instrument \cite{Optics11}. Furthermore, the average of these measurements at different locations on the chip along with their one-sigma standard deviation are depicted in Fig. \ref{indentFig} for each sensor. Additionally, a comparison of the Young's modulus measurements with the earlier studies in the literature is plotted in Fig. \ref{indentFig}. As seen in Fig.4. data shows
a similar trend to earlier work in literature. We attribute the differences in the reported Young’s modulus values to the differences in preparation and testing methods. We also note that there is a disparity in our measurements and a considerably higher mismatch for the mixing ratios 30 and 40 as shown in Fig.4.

\subsection{Effect of cladding material on the sensitivity}
\begin{figure}[!t]
\centering
\includegraphics{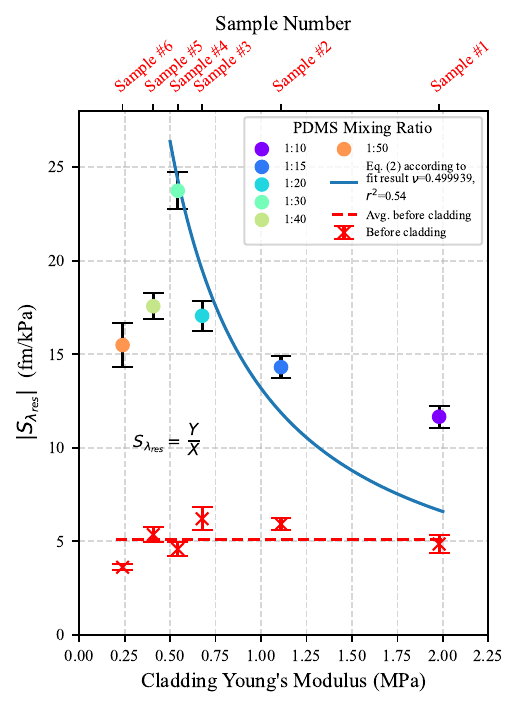}
\caption{Resonance wavelength sensitivity of the PDMS coated silicon photonic ultrasound sensors before and after PDMS coating. The stiffness of the cladding materials along the $x$-axis is measured with nano-indentation after the ultrasound characterization experiments near the sensors as depicted in Fig. \ref{indentFig}.}
\label{senstivityFig}
\end{figure}

\noindent The sensor sensitivity, denoted as $S_{\lambda_{res}}$, is expressed in a change in optical resonance wavelength per applied pressure, with units of $fm/kPa$. This sensitivity provides a measure of the waveguide's sensitivity since it does not depend on the linewidth of the photonic resonator or the read-out method that is used. The calculation of $S_{\lambda_{res}}$ from the measurements is explained in \ref{sec:sensor_def}. Fig. \ref{senstivityFig} illustrates the measured sensitivity of the sensors before and after PDMS coating, each sample featuring distinct cladding stiffness. We use Eq. \ref{eq1} to perform a curve fit by using the data for the PDMS cladded samples from number 1 to sample number 4. We assume $p_{12}$ value for the PDMS as $0.31$, which is the elasto-optic coefficient of Polystyrene \cite{Yariv_Yeh_1984}. We calculate $\frac{\lambda _r}{n_g}$ as \qty{376}{nm} and $\frac{\partial n_e}{\partial n_c}$ as 0.23 by using the resonator's dimensions and the mode simulations. The fit resulted in a Poisson's ratio of 0.499939 with a coefficient of variation, $r^2$, value of 0.54. Although the fitted curve does not follow precisely the measured values, it still provides us with an indicative measure of the sensitivity.

In the case of no-cladding, the sensor is fully submerged in the water, which is in contact with the silicon waveguide core and the silicon-dioxide bottom cladding. Hence, the top cladding material can be considered as water in the case of no cladding. When the top cladding of the waveguide is water, the effective refractive index modulation of the waveguide and the ultrasound sensing happens due to the two physical effects: the change of the refractive index of the water around the ring waveguide \cite{water_refractive,waxler_effect_1963}, and the change of refractive index of the silicon core and silicon-dioxide bottom cladding \cite{hazan_silicon-photonics_2022}. Here, we consider the deformations of the waveguide core structure as negligible since the Young's Modulus of the core and bottom cladding materials are around $170\ GPa$ and $70 \ GPa$, respectively. Although water can not deform, the longitudinal ultrasound pressure causes variation of the refractive index of the water. \cite{waxler_effect_1963,ZANELLI2006e105}.

The results in Fig. \ref{senstivityFig} indicate that optimizing the stiffness of the cladding can enhance sensitivity fourfold compared to sensors lacking a cladding layer and twofold compared to sensors with PDMS cladding at a saturated cross-link density. To our surprise, the results of the sensors with a Young's modulus lower than \qty{500}{kPa} disagree with the trend we observe with the sensors with a Young's modulus higher than \qty{500}{kPa}. We also note that for the stiffer formulations of PDMS cladding, $S_{\lambda_{res}}$ is higher than the anticipated values by the Eq. \ref{eq1}.  In the discussion section, we elaborate on these contradictory results.

\subsection{Noise Equivalent Pressure Density (NEPD) of the sensor}
\begin{figure*}[ht]
\centering
\includegraphics{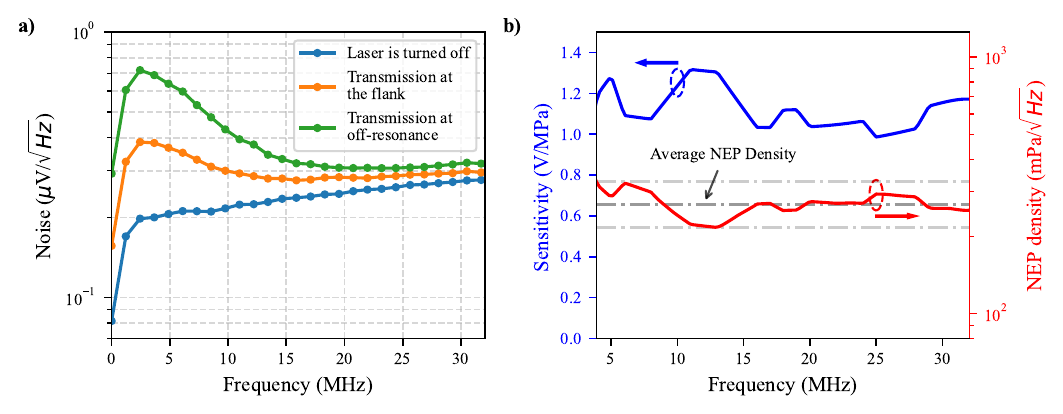}
\caption{Ultrasound response of the silicon photonic sensor with laser input power intensity of \qty{3}{\deci\belm} and the noise characterization. a) Power spectral density of the AC noise recordings from the sensor output of the sensor with the 1:30 mixing ratio. Flank and off-flank wavelengths are indicated in Fig. \ref{characterizationFig}(a). The Welch method is used to calculate the spectrums. No averaging and no ultrasound pulse were applied for the noise recordings. b) Calibrated sensitivity and noise equivalent pressure density spectrum of the sensor. Plotted for the frequency range where the signal-to-noise ratio is bigger than 10.}
\label{nepFig}
\end{figure*}

\noindent To assess the noise equivalent pressure density (NEPD) of the sensor, a comprehensive analysis is conducted involving the acquisition of noise data in the absence of ultrasound pulses and without any form of data averaging. The spectrogram depicted in Fig. \ref{nepFig}(a) illustrates the power spectral density of the recorded noise signals under three distinct conditions: at the flank wavelength, during off-resonance transmission, and when the laser is deactivated. The off-flank transmission has a higher noise than the flank transmission due to the increased transmitted intensity. Each measurement entails recording time traces spanning a duration of \qty{2}{\ms} from the AC output of the photodetector. Subsequently, the Welch method is applied to compute the power spectral density (PSD) across the frequency spectrum. In Figure \ref{nepFig}(a), the blue curve corresponds to the combined effect of photodetector dark current noise and electronic noise.

A laser input power of \qty{3}{\deci\belm} is employed to capture noise characteristics at the flank and off-resonance wavelength. The computed PSDs are depicted in Fig. \ref{nepFig}(a) with orange and green curves. This specific optical power setting is also employed to characterize ultrasound sensitivity. It should be noted that the choice of this input power level is carefully calibrated to fall within the linear operating range of the photodetector. This power level is determined while ensuring the output power is kept below the maximum linear input power specification of the photodetector to avoid any non-linear effects arising from the photodetector.

To compute the Noise Equivalent Pressure Density (NEPD) of the sensor, an approach consistent with the ultrasound characterization methods detailed in the preceding sections and in the methods section of this paper is employed. Note that identical ultrasound pulse parameters are employed for NEPD at higher laser power and for ultrasound characterization at lower laser power. 

To capture the NEPD characteristics, the laser power is set to \qty{3}{\deci\belm} while the wavelength remains fixed at the flank wavelength. Subsequently, the sensor's response to ultrasound pulses under these conditions is recorded. The calculated ultrasound sensitivity at this heightened laser power setting and the NEPD spectra of the sensor with a 1:30 PDMS
mixing ratio are depicted in Fig. \ref{nepFig}(b). The noise recorded at the flank wavelength and the sensitivity characteristic of the sensor are combined to determine the NEPD of the sensor as described in \cite{westerveld_sensitive_2021}. The sensor exhibits reasonably flat NEPD characteristics within the frequency range of \qty{3.88}{\mega\hertz} to \qty{32}{\mega\hertz} with an average NEPD of \qty[per-mode = fraction]{267}{\m\pascal\per\sqrt{Hz}}. In this analysis, we have chosen to disregard the frequency range where the signal-to-noise ratio of the measurement falls below 10. This decision is grounded to ensure an adequate level of pressure is applied to the sensor for accurately characterizing the NEPD. 

\section{Discussion}
\noindent 

The elasto-optic theoretical derivations (see Methods) and the experimental results presented in Fig. \ref{senstivityFig} indicate an inverse relationship between the stiffness of the cladding material and resonance wavelength sensitivity of silicon photonic ultrasound sensors. In previous studies, cladding the waveguide with PDMS has enhanced sensitivity \cite{hazan_silicon-photonics_2022}. 
Our results in Fig. \ref{senstivityFig} show that lower stiffness leads to a higher sensitivity, and therefore we conclude that the high sensitivity of PDMS can be attributed at least partly to its low stiffness compared to other cladding materials such as BCB, SU-$8$, or SiO$_2$.

It is of interest to determine the operation mechanism and effect of the cladding on the detected signal. Besides the effect of ultrasound pressure on the refractive index of the cladding material via the photoelastic effect, there is also the effect of the ultrasound pressure on the deformation and thickness of the cladding layer. The mode-field intensity distribution simulation in Fig. \ref{conceptFig}(d) shows that the optical field does not reach the interface between the cladding layer and the water, since the mode-field intensity drops to zero within a micrometer distance from the silicon photonic waveguide. This result suggests that sensors utilizing cladding material as the transduction mechanism only measure the changes in the refractive index of the cladding layer if the thickness of the cladding layer is sufficient enough. 

Using our theoretical derivation in the Methods section of this paper, we can express the refractive index of the cladding layer with linear opto-elasticity as

\begin{equation}
\label{eq2}
n_{PDMS}(P) = n_0+\frac{\partial n}{\partial \epsilon} \frac{1}{E}\frac{(1+\nu)(1-2\nu)}{1-\nu} P
\end{equation}
where $n_{PDMS}$ is the refractive index of the deformed PDMS under pressure, $n_0$ is the initial refractive index of the PDMS without pressure, $\epsilon$ is the out-of-plane strain in the cladding, and $P$ is the ultrasound pressure in the PDMS layer. Here we have assumed that the in-plane strain in the cladding is zero. Eq. (\ref{eq2}) suggests that a cladding material with a lower stiffness $E$ will lead to an increased change in refractive index. Nevertheless, the data in Fig. \ref{senstivityFig} does not exactly follow a $1/E$ dependence for all the samples. This might either be accounted for by changes in Poisson's ratio $\nu$ with varying mixing ratio, or by changes in the photoelastic properties that affect $\frac{\partial n}{\partial \epsilon}$. Eq. \eqref{eq2} shows that for materials which have a Poisson's ratio near $\nu=0.5$, one should actually aim to maximize $(1-2\nu)/E$ instead of $1/E$.

Results of our analysis and the direction provided by the elasto-optic theory in this paper, lead us to enhance the sensitivity of our sensors by a factor of two compared to the sensitivity of the sensor with a saturated cross-link density. However, we remark that the model is not free from imperfections. A further detailed study is required to elucidate the effect of the cross-linking density, Poisson's ratio, elasto-optic coefficient, Young's modulus and bulk modulus of the material on the sensitivity of polymer cladded silicon photonic ultrasound sensors. Thus the observations in this work point us toward further optimization of the cladding material. 
  
We compare the NEP density characteristics of the PDMS-cladded silicon photonic ultrasound sensors with the commercial piezoelectric transducers. We follow the method described in \cite{westerveld_sensitive_2021} and \cite{wissmeyer_looking_2018} for the comparison and use NEP density multiplied by the square root of the sensor area as the metric. The sensor area is an essential parameter for applications that require ultrasound sensor arrays placed at a small pitch, such as photoacoustic tomographic imaging \cite{hazan_silicon-photonics_2022}. This metric is estimated for the piezoelectric sensors as being in the range of 1.1-\qty{3.5}{mPa}\,\qty{}{\mm}\,\qty{}{\sqrt\hertz}. For example, needle hydrophones of various sizes have a range of 12.8-\qty{26.2}{mPa}\,\qty{}{\mm}\,\qty{}{\sqrt\hertz} \cite{westerveld_sensitive_2021}. The sensor with optimized cladding elasticity in this paper achieved \qty{2.62}{mPa}\,\qty{}{\mm}\,\qty{}{\sqrt\hertz}, which is comparable to the theoretical estimate of piezoelectric sensors even without optimization for the sensor's optical performance.

The Q-factor of the ring resonators does not affect the sensitivity, $S_{\lambda_{res}}$, of the ring resonators. However, it can affect the NEP density of integrated photonic ultrasound sensors. In this study, the resonators had a range of Q-factors from $5\times10^3$ to $15\times10^3$. Ring resonators with a higher Q-factor can be employed to obtain improved NEP density values in ultrasound sensing \cite{lee_theoretical_2023}.

\section{Methods}

\subsection{Ultrasound characterization setup}
\noindent The performance of the ultrasound sensors was assessed in a high-purity water tank. A piezoelectric transducer (Olympus V316-SM) with a center frequency of \qty{20}{\mega\hertz} and an element size of \qty{3.175}{\mm} was directly connected to an arbitrary waveform generator (Keysight Technologies 33522B, \qty{30}{\mega\hertz}) to produce ultrasound pulses. To interrogate the sensor, a tuneable laser (Keysight 81940A) which was incorporated into a multi-meter (Keysight 8164A), was used. Fiber optics with polarization control (Thorlabs FPC562) were used to direct the light to the silicon photonics ultrasound sensor, and polarization was set to the maximum transmission point for each experiment. The light was detected using a photodetector (Newport 1811-FC-AC) with two outputs: a DC output low-pass filtered at \qty{50}{\kilo\hertz} for measuring DC optical transmission and an AC output band pass filtered from \qty{25}{\kilo\hertz} to \qty{125}{\mega\hertz} for measuring the ultrasound signal. These outputs were digitized using an oscilloscope (RS RTA4004, \qty{200}{\mega\hertz} bandwidth) triggered by the arbitrary waveform generator, which generates an ultrasound pulse at each measurement trigger (Fig. \ref{setupFig}). Real-time ultrasonic time-of-flight measurements were performed to adjust the distance between the ultrasound source and the silicon photonic ultrasound sensor to \qty{38}{\mm}, corresponding to a target of \qty{26}{\micro\second} of arrival delay. The in-plane alignment was achieved by optimizing the signal amplitude.

\subsection{Pressure calibration measurements}
\noindent 
A calibrated needle hydrophone (PA LTD. - SN1302) with a diameter of \qty{0.075}{\mm} is used to calibrate the pressure measurements. The needle hydrophone is attached to a pre-amplifier and placed in a holder connected to a motorized 3D scanning system. The needle hydrophone and the ultrasound transducer  (Olympus – V316) are placed in a high-purity water tank. The ultrasound transducer is placed across the needle hydrophone at approximately \qty{38}{\mm}. A DC coupler (PA LTD.-SN: DC2000585) and a \qty{20}{\dB} high-frequency amplifier (HP-8447A) are attached to the needle hydrophone. The output of the amplifier is connected to an oscilloscope (Agilent DSO6034A) to digitize and record the received pressure pulses. The ultrasound transducer is connected to the arbitrary waveform generator (Keysight Technologies 33522B, \qty{30}{\mega\hertz}). Gaussian pulses with a bandwidth of \qty{8}{\mega\hertz} a center frequency of \qty{18.53}{\mega\hertz} with different peak-to-peak amplitudes are generated using the arbitrary waveform generator and applied to the ultrasound transducer. The distance between the needle hydrophone and ultrasound transducer is adjusted by monitoring the time delay of the pulse received by the oscilloscope so the generated ultrasound pulse hits the needle hydrophone at \qty{26}{\micro\second}. Following the distance alignment, an in-plane scan was performed by using the needle hydrophone at the same distance to find the maximum response-generating point in the measurement plane. Finally, Gaussian pulses in a peak-to-peak amplitude range of \numrange{0}{5}\unit{V} are generated, and the response of the needle hydrophone is recorded. The DC offset in the recorded responses of the needle hydrophone is removed, and \qty{20}{\dB} is subtracted to remove the amplification of the high-frequency amplifier. NIS traceable hydrophone calibration data from PA LTD. (ISO9001 certified laboratory) is used to calculate the peak pressure applied at the location of the silicon photonic sensors. These values are used for calculating the sensors' intensity sensitivity in Fig. \ref{characterizationFig}(d), and the NEP density in Fig. \ref{nepFig}(b).
\subsection{PDMS cladding fabrication}
\noindent In this study, Sylgrad 184 silicon elastomer (PDMS) is used as a cladding material for silicon photonic ultrasound sensors. PDMS is supplied in two parts from the manufacturer, a base (part A) and a cross-linking agent (part B). To prepare claddings of different cross-linking densities, the base and cross-linking agent are mixed thoroughly for \qty{5}{\minute} in six separate Petri dishes, each with a distinct mixing-ratio as reported in Fig. \ref{indentFig} and Fig. \ref{senstivityFig}. Subsequent to the mixing process, all uncured samples are placed in a vacuum desiccator for \qty{30}{\minute} to mitigate the formation of bubbles through degassing the mixture. The degassed samples are then spin-coated onto the silicon photonic microchips at a spin speed of \qty{6000}{\rperminute} for \qty{3}{\minute}. Following the spin-coating, the chips are placed in an oven, and PDMS claddings are cured for \qty{2}{\hour} at \qty{80}{\degreeCelsius}. The ultrasound characterization experiments are conducted once the samples have sufficiently cooled down to reach room temperature.

\subsection{Theory and elasto-optic model for polymer cladded silicon photonic ultrasound sensors}
\noindent Ultrasound detection sensitivity of a silicon photonic resonator can be evaluated by the sensitivity, $S_{\lambda_{res}}$, which quantifies the changes in resonance wavelength due to impinging acoustic pressure on the resonator. The shift in resonance wavelength of the resonator due to induced ultrasound is caused by two mechanisms: a change in the effective refractive index of the propagating mode due to the photo-elastic effect, and changes in the resonator's geometry and the geometry of its waveguide. We can express the resonance wavelength sensitivity of such resonators as 

\begin{equation}
\label{eq3}
{S_{{\lambda _{res}}}} = \frac{\partial \lambda _r}{\partial P} = \frac{\partial \lambda _r} {\partial {n_e}}  \frac{\partial {n_e}}{\partial P}
\end{equation}
where $\lambda_{r}$ is the resonance wavelength of the silicon photonic resonator used for detection. Additionally, when employing highly photo-elastic materials as cladding, it becomes possible to neglect the effective index change caused by waveguide geometry deformations \cite{hazan_silicon-photonics_2022}. Then, the predominant factor that influences the sensitivity is the alteration in the effective refractive index of the propagating mode.

To evaluate the sensitivity of a resonator, the first term on the right-hand-side of equation (\ref{eq3}) can be written as
\begin{equation}
\label{eq4}
\frac{\partial {\lambda _r}}{\partial {n_e}} = \frac{\lambda _r}{n_g}
\end{equation}
where $n_{g}$ is the waveguide group index calculated from the first-order dispersion of the propagating mode. This term corresponds to the design of the waveguide and the resonator. It can be calculated given the material properties together with the dimensions of the resonator by using a finite element method software. 

To understand the effect of material properties on the $S_{\lambda_{res}}$, we can further expand the second term on the right side of the (\ref{eq3}). It corresponds to a change in the effective refractive index of the propagating mode in the waveguide due to a change in pressure. The effective refractive index of a waveguide depends on the waveguide geometry and refractive index of the waveguide materials. In the case of a high-elasto-optic effect in the cladding material, we can express the second term as
\begin{equation}
\label{eq5}
\frac{\partial {n_e}}{\partial P} =\frac{\partial {n_e}}{\partial {n_c}}  \frac{\partial {n_c}}{\partial P}
\end{equation}
where $n_c$ is the material refractive index of the cladding. The first term on the right-hand side represents the optical mode field interaction with the cladding material. This term can be computed given the material refractive indices and waveguide geometry. In general, for the transverse magnetic-like mode, the interaction of the propagating light in the waveguide with the cladding material is expected to be higher than the transverse electric-like mode for the same geometry of a waveguide \cite{ravi_kumar_enhanced_2019,ding_broadband_2022}. Thus, using a TM mode waveguide can effectively improve this term. However, using a TM mode increases losses in bend sections of the waveguides. This ultimately increases the size of individual sensors and the minimum pitch of the sensor arrays or matrices.

The cladding material deforms upon impinging longitudinal pressure of the ultrasound wave, and this pressure field changes the molecular orientations in the material. The direction of the wave propagation is on the same axis as the particle motion. This motion creates uniaxial strain along the Y-axis (Fig. \ref{conceptFig}). For a uniaxial strain in the material, we can express the strain in each axis as $\epsilon_x=\epsilon_z=0$ and $\epsilon_y\neq0$. 

The motion of the molecules leads to a change in the material's refractive index both in the direction of the ultrasound wave (Y-axis) and in the transverse directions (X and Z). According to elasto-optic theory \cite{Yariv_Yeh_1984}, for small changes in the refractive index, the isotropic change in material refractive index in each direction can be written as
\begin{equation}
\label{eq6}
	\Delta {n_x} =  - \frac{1}{2}n_c^3{p_{12}}\epsilon_{y}
\end{equation}
\begin{equation}
\label{eq7}
	\Delta {n_y} =  - \frac{1}{2}n_c^3{p_{11}}\epsilon_{y}
\end{equation}
where $p_{11}$ and $p_{12}$ are the components of the elasto-optic tensor and $\epsilon_{y}$ is the strain in the material due to impinging pressure wave.

Additionally, by using the isotropic linear elastic deformation theory for small strains and considering a transverse electromagnetic TE-like mode propagation in the waveguide, the second term in (\ref{eq5}) can be expressed as 
\begin{equation}
\label{eq8}
\frac{{\partial {n_c}}}{{\partial P}} = \left[ { - \frac{1}{2}n_c^3{p_{12}}\frac{1}{E}\left( {\frac{{\left( {1 - 2\nu} \right)\left( {1 + \nu} \right)}}{{1 - \nu}}} \right)} \right]
\end{equation}
where $E$ and $\nu$ are Young’s Modulus (stiffness) and Poisson’s ratio of the cladding material, respectively.

Finally, using (\ref{eq3}), (\ref{eq4}), (\ref{eq5}), (\ref{eq8}), we can write: 
\begin{equation}
\label{eq9}
 	{S_{{\lambda _{res}}}} = \frac{{{\lambda _r}}}{{{n_g}}}  \frac{{\partial {n_e}}}{{\partial {n_c}}}  \left[ { - \frac{1}{2}n_c^3{p_{12}}\frac{1}{E}\left( {\frac{{\left( {1 - 2\nu} \right)\left( {1 + \nu} \right)}}{{1 - \nu}}} \right)} \right]
\end{equation}  

Equation (\ref{eq9}) suggests the sensitivity scales inversely with the cladding material’s stiffness.

\section{Conclusion}
\noindent In conclusion, we have demonstrated a route for improving the sensitivity of silicon photonic ultrasound sensors by adjusting the elasticity of the cladding polymer. Firstly, by decreasing the cladding polymer cross-linking density, the sensor achieved a four-fold increase in sensitivity compared to the sensors without top cladding. Moreover, it achieved two-fold higher resonance wavelength sensitivity to ultrasound pressure than the sensor with a saturated cladding polymer cross-linking density (Fig. \ref{senstivityFig}). Secondly, the sensor exhibited an average NEP density of \qty[per-mode = fraction]{267}{\m\pascal\per\sqrt{Hz}} across a wide frequency range of \qty{3.88}{\mega\hertz} to \qty{32}{\mega\hertz} when calibrated to a needle hydrophone (Fig. \ref{nepFig} (b)). The NEP density of this level with a small footprint performs comparable to the theoretical estimate of piezoelectric sensors and enables the silicon photonic ultrasound sensors for advanced imaging applications that require a small pitch of detectors.  Moreover, this study provides valuable insights into the influence of elasto-optical properties of cladding materials on the sensitivity of ultrasound sensors, which enables us to design tailored materials for the cladding of silicon photonic ultrasound detectors. 

Furthermore, enhancing the detection limit of the ultrasound sensors is crucial for photoacoustic imaging applications as it determines the trade-off between image resolution and imaging depth since the attenuation of acoustic waves in tissue correlates with the acoustic frequency and imaging depth. By enhancing the sensitivity of the photoacoustic imaging sensors, direct improvement of both the depth and resolution of the resulting images are possible. This can enable advanced applications of photoacoustic imaging, such as intra-vascular imaging with catheters or living mouse brain imaging. Additionally, the scalability of silicon photonics technology allows the fabrication of sensor arrays on a single chip with a small pitch without compromising sensitivity. Notably, the sensitivity enhancement demonstrated in this paper is compatible with the scalable CMOS processes. In fact, it merely necessitates an additional spin coating step, which is already available in CMOS fabrication centers; therefore, the cost-effective integration of these sensors into photoacoustic imaging probes is possible. With this promising demonstration, silicon photonic ultrasound sensors with tailored polymer claddings hold great potential to advance photoacoustic imaging applications.

\section{Acknowledgment}
\noindent 
The authors thank Rik Vos for providing the calibrated needle hydrophone and the calibration measurement setup.
\bibliographystyle{ieeetr}  
\bibliography{refs.bib} 
\end{document}